\documentclass[reprint, aps, prapplied, amsmath, amssymb, twocolumn, superscriptaddress, longbibliography]{revtex4-2}
\usepackage{xcolor}
\usepackage[normalem]{ulem}
\usepackage[utf8]{inputenc}
\usepackage{graphicx}
\usepackage{amsmath}
\usepackage{textgreek}
\usepackage{braket} 
\usepackage{enumitem}
\usepackage{indentfirst}
\usepackage[separate-uncertainty=true]{siunitx} 
\newcommand{\RNum}[1]{\uppercase\expandafter{\romannumeral #1\relax}}
\usepackage{titlesec}
\titlespacing*{\section}{0pt}{1\baselineskip}{0.5\baselineskip}
\titlespacing*{\subsection}{0pt}{0.5\baselineskip}{0.5\baselineskip}
\usepackage{verbatim}
\renewcommand{\selectlanguage}[1]{}

\begin{document}
\title{Role of Oxygen in Laser Induced Contamination at Diamond-Vacuum Interfaces}

\author{Shreyas Parthasarathy}
 \affiliation{Department of Physics, University of California, Santa Barbara, Santa Barbara, CA 93106, USA.\looseness=-1}

\author{Maxime Joos}
 \altaffiliation{Present address: Cailabs, 35000 Rennes, France.}
 \affiliation{Department of Physics, University of California, Santa Barbara, Santa Barbara, CA 93106, USA.\looseness=-1}

 \author{Lillian~B. Hughes}
 \affiliation{Materials Department, University of California, Santa Barbara, Santa Barbara, CA 93106, USA.\looseness=-1}
 
\author{Simon~A. Meynell}
 \affiliation{Department of Physics, University of California, Santa Barbara, Santa Barbara, CA 93106, USA.\looseness=-1}
 
\author{Taylor~A. Morrison}
 \affiliation{Department of Physics, University of California, Santa Barbara, Santa Barbara, CA 93106, USA.\looseness=-1}

\author{J.~D. Risner-Jamtgaard}
\affiliation{Stanford Nano Shared Facilities, Stanford University, Palo Alto, CA 94305, USA\looseness=-1}

\author{David~M. Weld}
 \affiliation{Department of Physics, University of California, Santa Barbara, Santa Barbara, CA 93106, USA.\looseness=-1}

\author{Kunal Mukherjee}
 \affiliation{Department of Materials Science and Engineering, Stanford University, Palo Alto, CA 94305, USA.\looseness=-1}

\author{Ania~C. Bleszynski Jayich}
 \email{ania@physics.ucsb.edu}
 \affiliation{Department of Physics, University of California, Santa Barbara, Santa Barbara, CA 93106, USA.\looseness=-1}
\date{January 12, 2024}

\begin{abstract}
Many modern-day quantum science experiments rely on high-fidelity measurement of fluorescent signals emitted by the quantum system under study. A pernicious issue encountered when such experiments are conducted near a material interface in vacuum is ``laser-induced contamination" (LIC): the gradual accretion of fluorescent contaminants on the surface where a laser is focused. Fluorescence from these contaminants can entirely drown out any signal from e.g. optically-probed color centers in the solid-state. Crucially, while LIC appears often in this context, it has not been systematically studied. In this work, we probe the onset and growth rate of LIC for a diamond nitrogen-vacancy center experiment in vacuum, and we correlate the contamination-induced fluorescence intensities to micron-scale physical build-up of contaminant on the diamond surface. Drawing upon similar phenomena previously studied in the space optics community, we use photo-catalyzed oxidation of contaminants as a mitigation strategy. We vary the residual oxygen pressure over 9 orders of magnitude and find that LIC growth is inhibited at near-atmospheric oxygen partial pressures, but the growth rate at lower oxygen pressure is non-monotonic. Finally, we discuss a model for the observed dependence of LIC growth rate on oxygen content and propose methods to extend \textit{in situ} mitigation of LIC to a wider range of operating pressures.
\end{abstract}
\maketitle
\section{Introduction}

Modern quantum science experiments often deploy a combination of vacuum environments, engineered material interfaces, and high-intensity light to achieve the degree of control and stability necessary for sensing, computation, and simulation. Vacuum plays the dual role of enabling cryogenic operation and mitigating harmful contaminants on material surfaces. These surfaces can limit qubit performance in a variety of ways: adsorbates on electrode surfaces generate electric field noise in ion traps \cite{chiaverini_insensitivity_2014,kim_electric-field_2017,brown_materials_2021}; charge traps in diamond limit stability of near-surface nitrogen-vacancy (NV) centers \cite{bluvstein_identifying_2019,chu_coherent_2014}, surface-related two-level systems limit superconducting qubit scalability and optomechanical resonator lifetimes \cite{gao_semiempirical_2008,crowley_disentangling_2023,maccabe_nano-acoustic_2020}. While vacuum and low temperature alone sometimes ameliorate these issues and enable access to new physical regimes, the introduction of high-intensity light can counter-productively generate surface contamination that shortens experimental lifespans. Here, we characterize this ``laser-induced contamination" (LIC) -- the progressive accumulation of photo-chemically bonded, fluorescent matter at the site of a laser focused onto a surface -- and its effect on visible wavelength color center experiments. 

The NV center in diamond is an optically addressable qubit known for its stability in ambient conditions. Despite the ostensible convenience of atmospheric conditions, vacuum and cryogenic operation of NV centers provide many advantages, and in some cases is required. A variety of condensed matter phenomena that are targets of sensing experiments only manifest at low temperatures \cite{jenkins_imaging_2022, borst_observation_2023}. Molecular targets (e.g. spin labels and biomolecules) can be susceptible to photo-damage that is exacerbated at elevated temperatures or pressures  \cite{pinto_readout_2020,eckardt_stability_2015}. Low-temperature operation also facilitates spin-photon entanglement \cite{bernien_heralded_2013}, single-shot readout \cite{irber_robust_2021}, and decreased spin relaxation and decoherence rates \cite{takahashi_quenching_2008, jarmola_temperature-_2012, bar-gill_solid-state_2013}. Additionally, improving spin coherence and charge stability of the shallow NV's used in highly sensitive nanoscale magnetometry experiments is an outstanding challenge that should benefit from the advanced level of surface characterization (e.g. LEED, XPS, NEXAFS) \cite{dontschuk_x-ray_2023, kawai_nitrogen-terminated_2019}, treatment (e.g. annealing, surface cleaning) \cite{sangtawesin_origins_2019, kim_effect_2014}, and control (over e.g. surface adsorbates \cite{neethirajan_controlled_2023, hauf_chemical_2011, zvi_engineering_2023}) that is accessible in an ultra-high vacuum experiment. Fluorescent LIC is a persistent barrier to such progress; it has been observed in many vacuum NV experiments, but to our knowledge is not yet discussed in the literature. 

\begin{figure}[h]
    \includegraphics[width=3.375in]{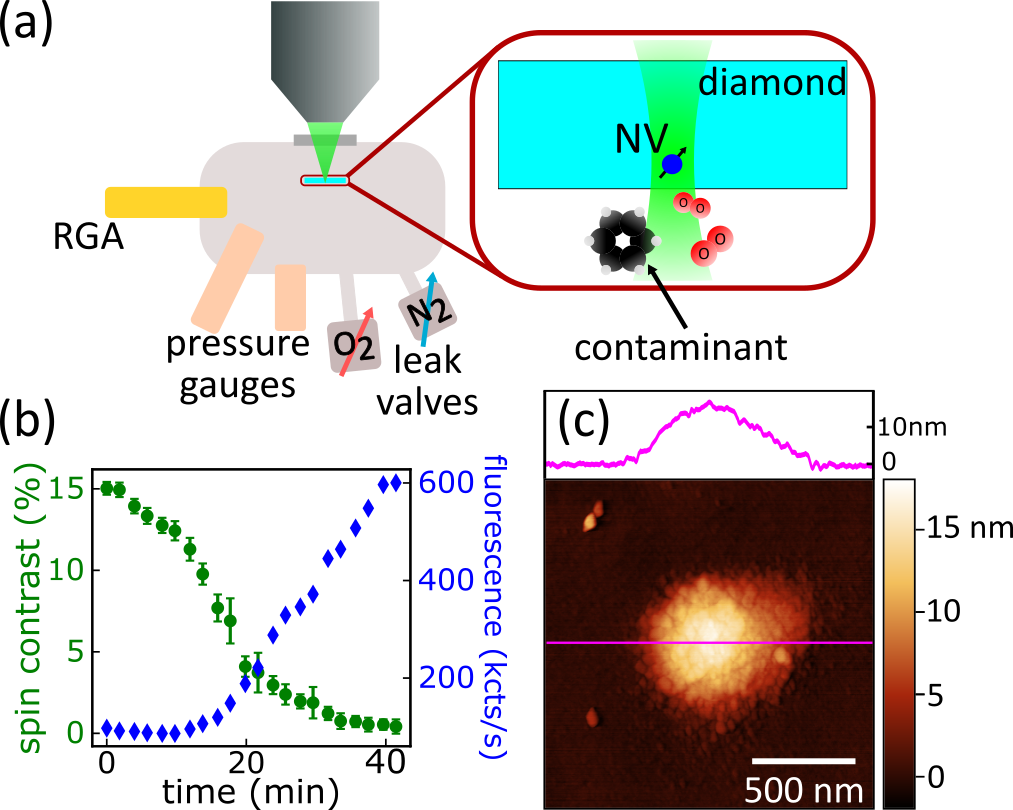}
    \caption{(a) Schematic of vacuum chamber and experiment. A diamond sample with near-surface NV centers resides inside a vacuum chamber outfitted with oxygen and nitrogen leak valves, a residual gas analyzer (RGA), and pressure gauges. An ambient microscope objective transmits 532 nm laser excitation and collects red fluorescence from both NVs and LIC.
    Inset shows a schematic view of the photochemistry underlying LIC: organic molecules, oxygen, and the diamond surface react under intense light immediately adjacent to the NV under study. (b) Total fluorescence and NV spin contrast plotted as a function of exposure time to laser illumination in vacuum. LIC-induced fluorescence eventually drowns out NV fluorescence, reducing spin contrast to near zero. 
    (c) Atomic force microscopy image and linecut (top) of laser-induced contamination grown on bulk diamond 
    after 2 days of nearly continuous illumination. }
    \label{fig:schematic}
\end{figure}

An example of LIC and its adverse impacts on NV center experiments is depicted in Fig.~\ref{fig:schematic}. Here, LIC forms when a green laser is focused on a diamond surface held in the vacuum chamber depicted in Fig.~\ref{fig:schematic}(a). As LIC-induced background fluorescence accrues over time, it masks the spin-state dependent fluorescence of the NV center (Fig.~\ref{fig:schematic}(b)), ultimately limiting the duration of the experiment. Moreover, the physical accumulation of contaminants at the site of measurement (characterized using atomic force microscopy (AFM) in Fig.~\ref{fig:schematic}(c) confounds study of other surface-related phenomena \cite{kim_decoherence_2015,sangtawesin_origins_2019} and hinders experiments which require proximity to a target, such as scanning NV magnetometry \cite{maletinsky_robust_2012}. 

Though not actively studied in the solid-state qubit community, LIC has been extensively characterized in the context of optics in space (primarily in the ultraviolet \cite{hatheway_contamination_2015, wagner_laser_2020, gebrayel_el_reaidy_study_2018, tighe_growth_2008, riede_laser-induced_2011} and near-infrared \cite{bartels_removal_2019, ling_impact_2009, hovis_optical_1994} regimes). In the vacuum of space, LIC has the harmful capacity to reduce transmission through optical components and introduce wavefront error, threatening the commissioned lifetime of many satellite-based experiments where routine cleaning cannot occur (e.g. gravitational wave observation \cite{bartels_laser-induced_2023, exarhos_optical_2013}, LIDAR-based surveys of planets/moons \cite{chen_contamination_2006}, atmospheric dynamics and aerosol measurement \cite{hovis_qualification_2006, wernham_laser-induced_2010}, etc. Summary in \cite{riede_laser-induced_2011}). This research community has identified hydrocarbon precursors for LIC and developed mitigation strategies tailored primarily to silica \cite{tighe_growth_2008, egges_laser-induced_2016, wernham_laser-induced_2010}, often exploiting the well-established capacity for reactive oxygen species to degrade hydrocarbon contaminants \cite{anglada_interconnection_2015}.

In this work, we use the chamber depicted in Fig.~\ref{fig:schematic} to systematically explore the conditions for LIC growth in an NV center experiment. Inspired by progress on mitigating LIC in the aforementioned UV/IR space-based experiments, we expose a diamond surface to varying levels of oxygen and measure the resulting LIC growth under green laser illumination. Here, we report four central observations which we believe pertain to all optically addressable solid-state systems (e.g. see Ref. \cite{sup_note} for LIC on silicon). First, we report significant LIC-induced fluorescence on diamond at room temperature for oxygen pressures ranging from $10^{-8}$ mbar to $10^{-2}$ mbar. Second, LIC fluorescence levels and physical volumes of LIC are correlated, in line with studies of LIC on other surfaces \cite{schroder_fluorescence_2007, egges_laser-induced_2016, wernham_laser-induced_2010}. Third, the accumulation of LIC during NV measurements is reproducibly controlled by residual oxygen concentration. This accumulation rate depends non-monotonically on oxygen content: LIC growth is slow in a low oxygen environment, dramatically faster at intermediate oxygen pressures, and fully inhibited at near-atmospheric pressures (consistent with routine LIC-free operation in ambient conditions). Fourth, we demonstrate that illumination under oxygen-rich environments can be used as an \textit{in-situ} method to remove LIC in systems where contamination is unavoidable. We end by conjecturing possible mechanisms for LIC consistent with its ubiquity and observed kinetics and pathways for more efficient mitigation.

\section{Experimental Setup}
An electronic grade (100) diamond substrate (Element Six) with an epitaxial layer of $\sim \SI{100}{nm}$ of CVD-grown, isotopically purified (99.998\% $^{12}$C) diamond was implanted with $^{14}$N and oxygen-terminated using standard preparation procedures for shallow NV center studies \cite{sup_note}. While LIC grows on the bulk diamond surface (Fig.~\ref{fig:schematic}(c)), we primarily used fabricated $\SI{670}{nm}$ diameter nanopillars to achieve more efficient excitation and fluorescence collection for NVs and LIC alike. 

Measurements were performed in a stainless steel vacuum chamber (Fig.~\ref{fig:schematic}(a)) with a water-dominated background pressure of $\SI{7e-8}{mbar}$, and sample temperature was $\SI{298}{K}$. To control oxygen content, 99.999\% molecular oxygen (Airgas OXR33A) was admitted into the chamber via an all-metal leak valve (Vacgen LVM) with electropolished stainless steel lines (Superlok USA). Oxygen partial pressure was measured using a residual gas analyzer (SRS RGA200) at low pressures, and multiple pressure gauges at higher pressures.

A full material inventory of the chamber is presented in Ref \cite{sup_note}. All materials were chosen to conform to low outgassing standards (ASTM E595-77), as is typical practice for many vacuum experiments. Some discussion of potential sources of contaminant precursors is contained in Ref. \cite{sup_note}. 

\section{Characterizing LIC Formation Rate}
\begin{figure}
\centering
\includegraphics[width=3.375in]{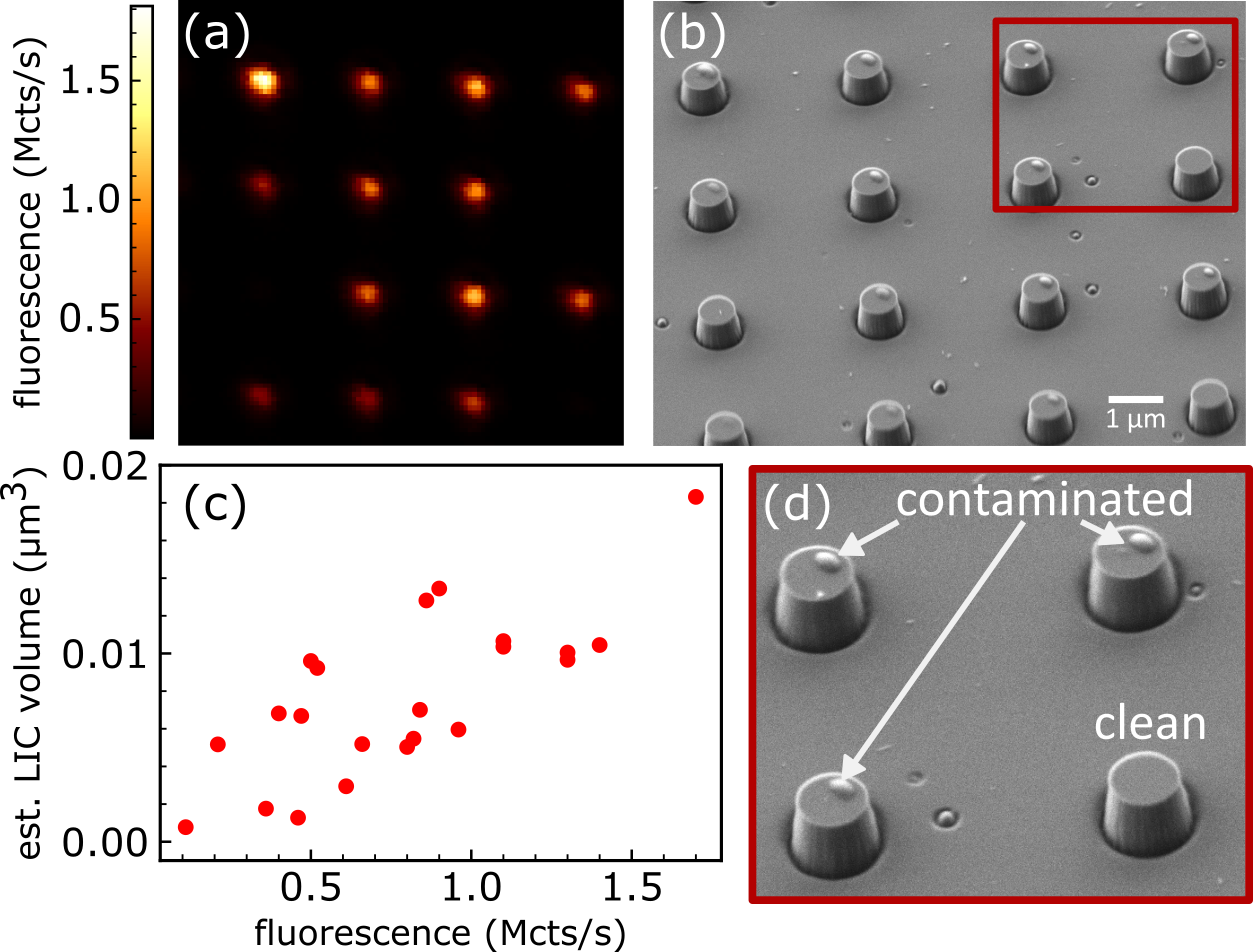}
\caption{(a) A confocal fluorescence image of LIC-contaminated nanopillars, taken at a low ($\sim$\SI{90}{\mu W}) laser power to minimize additional contamination during the scan. Comparison to standard NV fluorescence suggests counts are almost entirely ($>95\%$) due to LIC. (b) SEM image of the pillars shown in (a) after removal of sample from chamber. Contamination is visible on pillar tops, likely forming in the region of highest field intensity (off-center in this case \cite{sup_note}). (c) Scatter plot of estimated contaminant volume versus measured pillar fluorescence. A clear correlation is apparent. (d) A zoomed-in region of the SEM image in (b). The bottom right pillar, which was not contaminated, is shown for comparison.}
    \label{fig:contaminationSEM}
\end{figure}

Our \textit{in situ} probe of LIC was a home-built confocal microscope with $\SI{532}{\ nm}$ excitation and $652-1042 \SI{ }{\ nm}$ collection using a 0.7 NA objective. We focused 1.5 mW of laser power ($\sim\SI{220}{\ kW/cm^2}$) quasi-continuously on near-surface NVs in a series of nanopillars and tracked the build-up of fluorescence over time. Fig.~\ref{fig:contaminationSEM} compares characterization via confocal fluorescence measurements (under a lower excitation power of 92 $\mathrm{\mu}$W) and via \textit{ex situ} scanning electron microscopy (SEM). Fig.~\ref{fig:contaminationSEM}(a) is a fluorescence image of 16 nanopillars contaminated under varying illumination times (3-12 hours) and oxygen pressures ($10^{-4}$ to $10^{-2} \mathrm{\ mbar}$). After these exposures, LIC dominates the observed fluorescence. For comparison, typical NV fluorescence rates for the excitation power used in Fig.~\ref{fig:contaminationSEM}(a) are $\sim 50 \mathrm{\ kcts/s}$. We note hours-scale LIC-induced fluorescence growth is not a peculiarity of this setup: the example depicted in Fig.~\ref{fig:schematic}(b) was produced with a different chamber and sample, yet shows similar growth dynamics. 

Fig.~\ref{fig:contaminationSEM}(b) shows the same pillars as Fig.~\ref{fig:contaminationSEM}(a), imaged via SEM after removal from the vacuum chamber. A bright mound of contamination is visible on each nanopillar, from which contaminant volume can be estimated. Fig.~\ref{fig:contaminationSEM}(c) is a scatter plot of measured fluorescence and LIC volume estimated from the SEM image for 22 different pillars. The plot reveals a correlation between fluorescence intensity and the physical size of deposits. From this trend, we extract an LIC volume-to-fluorescence ratio for our setup of $\sim6\times 10^{-4} \mathrm{\mu m^3 s/Mcts}$ at $1.5$ mW. With this calibration, we use LIC fluorescence intensity growth as a proxy for volume growth (see Fig.~\ref{fig:ppo2}), a technique also deployed by the broader community \cite{schroder_fluorescence_2007, gebrayel_el_reaidy_study_2018, wernham_laser-induced_2010}).

\section{Role of Oxygen in Contamination Rate}
\begin{figure*}[htbp]
    \centering
    \includegraphics[width=6.75in]{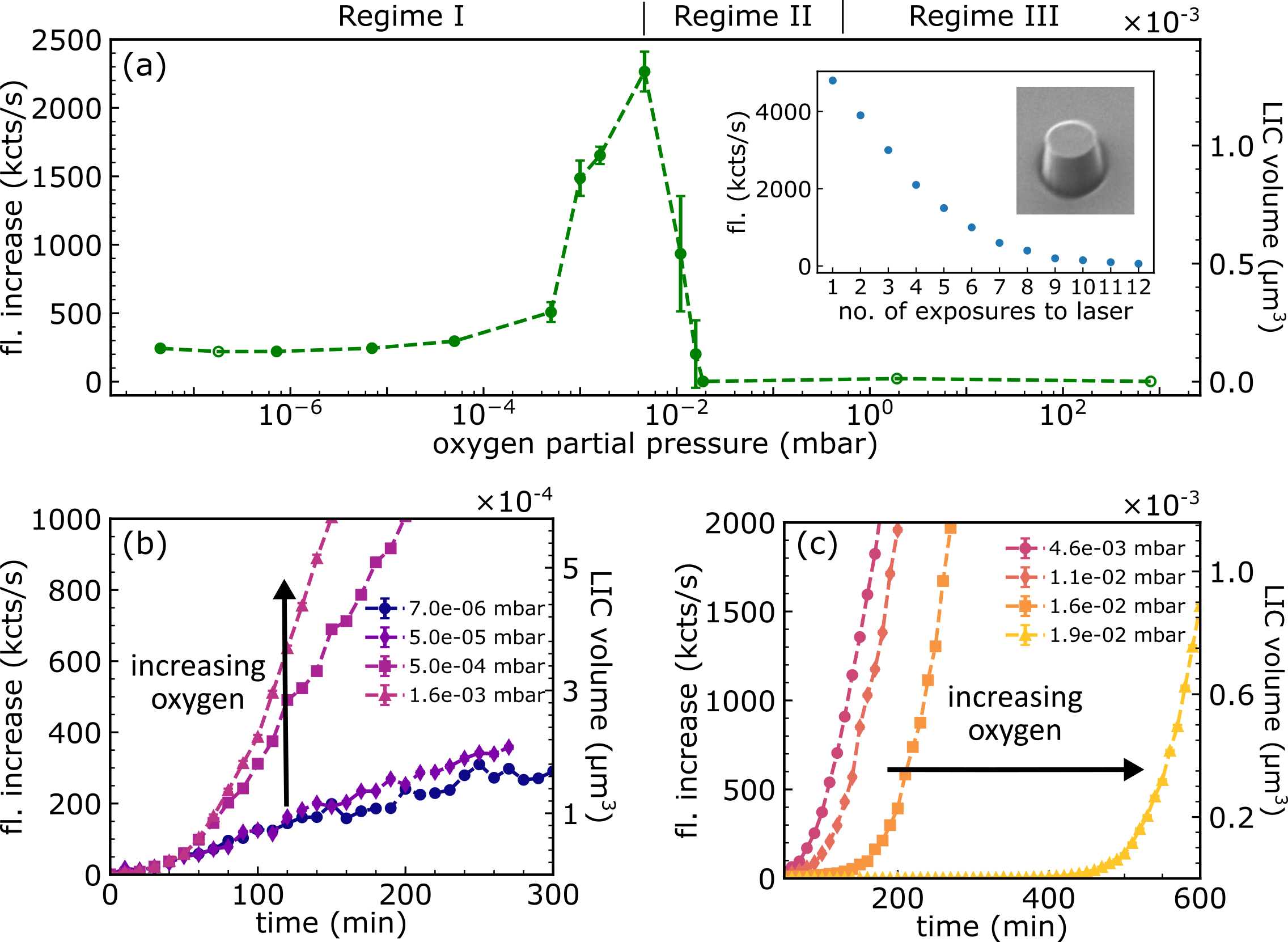}
    \caption{(a) Average change in fluorescence (fl.) and calculated LIC volume (using the correlation from Fig.\ref{fig:contaminationSEM}) after 200 minutes of laser illumination at various oxygen partial pressures. Each filled marker is the mean fluorescence increase averaged over multiple nanopillars contaminated at the same partial pressure. Error bars are 68\% confidence intervals. Unfilled markers are derived from observations of single nanopillars. Inset shows the degradation of LIC under successive exposures to light in an oxygen-rich (Regime \RNum{3}) environment.
    (b) Representative LIC growth curves for Regime \RNum{1}. Increasing oxygen pressure increases the growth rate. Error bars are smaller than the size of data points. Details of how fluorescence was sampled, averaged, and compared are in Ref. \cite{sup_note}.  
    (c) Representative LIC-induced fluorescence growth curves for Regime \RNum{2}. Increasing oxygen delays the onset of LIC growth.}
    \label{fig:ppo2}
\end{figure*}

We next examine the dependence of LIC formation on molecular oxygen partial pressure. Fig.~\ref{fig:ppo2}(a) plots the increase in fluorescence after 200 minutes of illumination at constant $1.5 \mathrm{ mW}$ as a function of measured partial pressure $P_{O_2}$. A control experiment using nitrogen gas confirms that the effect is due to only total oxygen content, rather than pressure-related effects alone \cite{sup_note}. We identify 3 different LIC formation regimes.

\emph{Regime \RNum{1}:} For $P_{O_2} \lesssim 5 \times10^{-3}$ mbar, the LIC growth rate \textit{increases} as $P_{O_2}$ increases. Fig.~\ref{fig:ppo2}(b) shows four representative curves of fluorescence as a function of illumination time. The onset of LIC growth is immediate, but the rate increases with increasing $P_{O_2}$, until growth is fastest at $P_{O_2} \sim 4.6 \times10^{-3}$ mbar.  

\emph{Regime \RNum{2}:} A distinctly different regime occurs for $2\times10^{-3} < P_{O_2} < 2$ mbar. Fig.~\ref{fig:ppo2}(c) plots representative fluorescence growth curves in this regime. Once contamination starts, the rate of LIC growth is similar to the peak growth rate observed in Regime \RNum{1}. However, onset of contamination is delayed, with a delay time that \textit{increases} as $P_{O_2}$ increases. In Fig.~\ref{fig:ppo2}(a), this effect manifests as an abrupt decrease in the average level of fluorescence after 200 minutes.

\emph{Regime \RNum{3}:} Finally, LIC ceases to form for $P_{O_2} > 2\mathrm{ mbar}$. In this regime, LIC is fully inhibited. Additionally, exposing existing LIC deposits to light removes their associated fluorescence. Subsequent SEM imaging of a formerly contaminated pillar treated with light confirms the physical removal of contaminants (see inset in Fig.~\ref{fig:ppo2}(a)).

\section{Model for Contamination}
Our results  provide valuable insight into the processes and reactants involved in the formation of LIC. The strongly non-monotonic dependence of LIC growth rate on oxygen content suggests that photoactivated molecular oxygen must participate in competing growth and etch processes on the diamond surface. At high oxygen content, etching of LIC is consistent with photo-catalyzed oxidation of an organic or graphitic contaminant. Carbon-based composition is consistent with reported spectroscopy of macroscopic LIC in other systems \cite{tighe_growth_2008} and our Auger-electron spectroscopy measurements \cite{sup_note}. The threshold pressure for oxidation is consistent with results for UV-catalyzed LIC on vacuum windows \cite{wernham_laser-induced_2010} suggesting inhibition occurs via gaseous oxidation independent of substrate. At lower oxygen pressures, oxidation is not the dominant process. Many other processes involving a variety of reactants (e.g. oxygen, organic LIC precursor molecules, water, dangling bonds on the diamond surface, etc.) are plausible.

Here we outline one model of competing reactions involving oxygen, hydrocarbons, and surfaces (in the spirit of Refs. \cite{stewart_absolute_1989, delzenne_photosensitized_1964, bhanu_role_1991}), and show that our model qualitatively reproduces the behavior depicted in Fig.~\ref{fig:ppo2}. Rate equations, numerics, and potential variants of the model can be found in Ref. \cite{sup_note}.

Our model supposes the LIC process consists of three separate reactions with first-order kinetics, all catalyzed by the photo-activation of reactants: 
\begin{enumerate}[label=\Alph*)]
    \item A surface-limited growth reaction at the vacuum-diamond interface between adsorbed oxygen and adsorbed LIC precusors (rate-limiting at early times and/or low $P_{O_2}$), obeying Langmuir-Hinschelwood kinetics \cite{atkins_atkins_2022}.
    \item An oxygen-independent reaction at the vacuum-LIC interface where gaseous LIC precursors attach to LIC deposits directly. This reaction dominates once a seed contaminant has a large enough surface area.
    \item An oxidation reaction that consists of photo-catalyzed reactive oxygen decomposing LIC.
\end{enumerate} 
 
We now discuss how this model can qualitatively produce the regimes in Fig.~\ref{fig:ppo2}:

\emph{Regime \RNum{1}:} Dynamics in Fig.~\ref{fig:ppo2}(b) are primarily driven by reaction $A$: as $P_{O_2}$ is increased, more oxygen adsorbs onto the diamond surface and encourages LIC growth via reaction $A$. At late times, all surface-mediated sites where reaction $A$ can proceed are depleted. Reaction $B$ is independent of adsorbates and thus dominates. Fast, linear growth occurs (see late times in Fig.~\ref{fig:ppo2}(b), $P_{O_2} = 1.6\times 10^{-3}\mathrm{mbar}$ or Fig \ref{fig:ppo2}(c)). 

\emph{Regime \RNum{2}:} For $P_{O_2} > 4.6 \times 10^{-3} \mathrm{mbar}$, two effects subdue the initial growth rate. First, adsorbed oxygen displaces adsorbed LIC precursors, which impedes growth via reaction $A$. Second, gaseous oxygen starts etching existing LIC (via reaction $C$), slowing net growth. Still, when enough LIC eventually forms at late times, reaction $B$ dominates and the same maximal growth rate is observed. Rather than complete elimination of LIC, inhibition first manifests as the delayed onset seen in Fig \ref{fig:ppo2}(c). 

\emph{Regime \RNum{3}:} Finally, when $P_{O_2}$ is high enough, reaction $C$ dominates and LIC ceases to form. 

\section{Mitigation Methods}
\subsection{Prevention}
Our measurements indicate that standard vacuum practices do not necessarily avoid LIC. Hence, in addition to undergoing species-insensitive outgassing tests, vacuum materials should be screened for chemical composition and/or compatibility in an LIC test chamber (as is already done in experiments with particularly stringent reliability requirements such as LISA, LIGO, etc. \cite{hatheway_contamination_2015, li_optical_1999, exarhos_laser-induced_2013, chen_contamination_2006}). To theoretically predict which materials are prone to producing LIC, direct chemical analysis of composition and formation is necessary. While the small volumes in this experiment are not resolvable via many \textit{in situ} characterization methods, additional analysis of the LIC absorption and fluorescence (e.g. observing spectral shifts as LIC grows or kinetics as a function of wavelength) could further elucidate reaction mechanisms (e.g. polymerization).
\subsection{Oxidation}
For NV experiments, venting with oxygen and removing LIC with the same laser used for spin readout is considerably less invasive than methods that require sample removal and \textit{ex situ} cleaning. This study suggests boosting the reactivity of residual oxygen in vacuum is key to inhibition of LIC at lower pressures. To that end, introducing atomic or UV/ozone-based oxygen treatments may increase the effective etch rate by many orders of magnitude. \cite{bartels_removal_2019}. Alternatively, ultra-high vacuum chambers may incorporate load-locks for the explicit purpose of periodic cleaning with oxygen and light without partial venting.

\subsection{Substrate Control}
A parallel substrate-specific route towards inhibition requires more detailed study of LIC growth rate as a function of surface preparation. Surface roughness, susceptibility to charging, baking to remove adsorbates, changing surface termination, and protective capping layers likely affect the kinetics of LIC formation (and in some cases, have already been shown to do so. See Supplemental \cite{sup_note} and elsewhere \cite{brown_physical_2019, riede_laser-induced_2011, wernham_laser-induced_2010, brown_materials_2021}). Finally, we note that while LIC formation reactions are photo-catalyzed (and thus temperature is not a primary activation mechanism), a small Arrhenius coefficient may still have a large and beneficial effect when anticipating the likelihood of LIC for cryogenic experiments.

\vspace{.1in}
\section{Conclusion}

In conclusion, we have shown the phenomenon of laser-induced fluorescence often seen in vacuum quantum science experiments is due to the photo-catalyzed accretion of physical material, a process also observed in space optics experiments. The contamination growth rate varies widely and non-monotonically with oxygen concentration; oxygen can either encourage or inhibit contamination due to multiple competing processes, but is an effective mitigant at high enough pressures. We propose one set of competing reactions involving the diamond surface, contaminants, adsorbed oxygen, and gaseous oxygen that qualitatively reproduces the behavior we observe. 

As lasers and vacuum become an indispensable element in increasingly complicated quantum science experiments, deep technical understanding of failure mechanisms is essential. Our approach suggests several pathways towards elimination of one particularly destructive mechanism. Moreover, light-induced surface deterioration in diamond, especially in vacuum, is known to limit shallow NV spin coherence and charge stability. While these changes are not well understood, we suspect the origin of these effects may be closely related to early-time LIC-related processes. Controlled experiments probing this chemistry will therefore be critical to enabling both LIC-free operation in vacuum and tackling these more subtle, yet fundamental limitations to defect centers in the solid state.

\begin{acknowledgments}
We gratefully acknowledge support of the U.S. Department of Energy BES grant No. DE-SC0019241.
We acknowledge the use of shared facilities of the UCSB Quantum Foundry through Q-AMASE-i program (NSF DMR-1906325), the UCSB MRSEC (NSF DMR 1720256), and the Quantum Structures Facility within the UCSB California NanoSystems Institute. D.M.W. and A.B.J. acknowledge support from the NSF QLCI program through grant number OMA-2016245. Portions of this work were performed in the UCSB Nanofabrication Facility, an open access laboratory, and at the Stanford Nano Shared Facilities (SNSF), supported by the National Science Foundation under award ECCS-2026822.
S.P. acknowledges support from the Department of Defense (DoD) through the National Defense Science and Engineering Graduate (NDSEG) Fellowship Program.  
L. B. H. acknowledges support from the NSF Graduate Research Fellowship Program (DGE 2139319) and the UCSB Quantum Foundry.
S. A. M. acknowledges support from the UCSB Quantum Foundry.
We thank Rob Knowles and Kurt Olsson for helpful discussions, and Pooja Reddy and Rachel Schoeppner for preliminary experiments.
\end{acknowledgments}

\bibliography{LIC_Paper.bib}
\nocite{hughes_two-dimensional_2023}
\nocite{mathieu_auger_1997}
\nocite{smith_surface_1994}
\nocite{kk_rohatgi-mukherjee_fundamentals_1986}
\nocite{tatsukami_reaction_1980}
\nocite{hansen_kinetics_1964}
\nocite{asua_polymer_2008}
\nocite{farneth_mechanism_1985}
\nocite{thomas_thermal_1992}

\end{document}


\title{Supplemental: Role of Oxygen in Laser Induced Contamination of Diamond-Vacuum Interfaces}

\author{Shreyas Parthasarathy}
\author{Maxime Joos}
\author{Lillian~B. Hughes}
\author{Simon~A. Meynell}
\author{Taylor~A. Morrison}
\author{J.~D. Risner-Jamtgaard}
\author{David~M. Weld}
\author{Kunal Mukherjee}
\author{Ania~C. Bleszynski Jayich}

\date{January 12, 2024}
\maketitle
\section{Additional Setup Details}
\subsection{Sample preparation}
The diamond substrate used in this work was an Element Six (100) electronic grade substrate of dimensions 4mm $\times$ 4mm $\times$ 0.5mm with the following history prior to the measurements in this paper:
\begin{enumerate}
    \item Polished to a thickness of $\sim 130 \mu$m (Syntek, Ltd.)
    \item $\SI{4}{\mu}$m strain relief etch in an \ch{Ar}/\ch{Cl2} plasma followed by a 1:1 mixture of \ch{H2 SO4} and \ch{H NO3}
    \item 100 nm of isotopically purified ($99.998\% \ch{^{12}C}$) PE-CVD grown diamond on top \cite{hughes_two-dimensional_2023}
    \item $^{14}\mathrm{N}$ implantation with 4 keV energy (7 nm expected depth), $\SI{2e10}{cm^{-2}}$ dose at a 7 degree tilt (Innovion Corp.).
    \item \SI{850}{C} vacuum anneal to reduce implantation damage and promote formation of NV centers
    \item Nanopillar lithography: A protective mask of 5 nm of \ch{Cr} was deposited onto the diamond followed by HSQ resist. Nanopillars were patterned via e-beam lithography. Diamond was etched in a \ch{Cl2}/\ch{O2} plasma (to remove \ch{Cr} mask) and then a \ch{Ar}/\ch{O2} plasma. The resist was then cleaned in buffered \ch{HF} followed by Chromium Etchant 1020.
    \item Sample was oxygen terminated via cleaning in a 1:1:1 mixture of \ch{H Cl O4}, \ch{H NO3}, and \ch{H2 SO4} and a subsequent anneal in an air oven at \SI{450}{C} for 4 hours.
    \item A platinum microwave stripline was deposited on the sample using standard contact lithography, with resist lift-off performed in N-methyl-2-pyrrolidone followed by acetone and isopropanol. 
    \item The sample was cleaned in a 1:1 mixture of \ch{HNO3} and \ch{H2SO4}, and then re-oxygen annealed in an air oven at \SI{450}{C} for 4 hours.
\end{enumerate}

\subsection{Fluorescence measurements}
For technical reasons, tracking fluorescence changes under truly continuous laser illumination was infeasible. Therefore, all measurements took place under ``nearly continuous" illumination, i.e. laser illumination with a duty cycle $\sim 95\%$.  We do not expect this difference to affect any of the results, and prior work with LIC in other systems suggests that ``mound"-like morphology is a signature of continuous wave illumination \cite{riede_laser-induced_2011}.

Each curve in Figure 3 in the main text was constructed as follows:
\begin{enumerate}
    \item For reasons related to NV measurements, each fluorescence measurement consisted of averaging the number of photons collected in thousands of collection windows each 350 ns in length, all under 1.5 mW illumination.
    \item Many measurements of fluorescence were taken in sub-two-minute intervals over the entire $>200$ minute illumination period
    \item The finely ($\approx 2$-minute) spaced data were interpolated to extract fluorescence over time at exact ten-minute markers, which simplified comparisons of different trials.
    \item The initial fluorescence was subtracted from the data so that only the fluorescence increase was measured.
    \item These regularly spaced data constitute the individual curves in Figs.~3(b) and 3(c).
    \item Comparing data points at the calculated 200-minute mark yields the data in Fig.~3(a). As is mentioned in the main text, each data point in Fig.~3(a) is an average across all nanopillars contaminated at the same oxygen partial pressure, with fluorescence vs time collected as described above. Error bars in Fig.~3(a) for each average of $n$ nanopillars were constructed using the standard error for a student's t distribution with $n-1$ degrees of freedom.
\end{enumerate}
\subsection{Vacuum chambers used in this work}
\label{sec:chambers}
Two different vacuum chambers were used in this work, here denoted as Chamber 1 and Chamber 2.

Chamber 1 is the chamber described in the main text and is custom-designed for ultra-high vacuum experiments, with the full contents itemized in the next section. Measurements used to produce Figs. 1(c), 2, 3 in the main text and Figs. \ref{supfig:n2} and \ref{supfig:power} were conducted using Chamber 1.

Chamber 2 is an Instec HCP421 vacuum chamber. Measurements used to produce Fig 1(b) in the main text and Fig. \ref{fig:silicon} were conducted using Chamber 2. All measurements in this chamber were conducted at a temperature of \SI{295}{K} and pressure of $\sim 5\times 10^{-6}$ mbar, with no characterization of residual vacuum content or screening of materials placed in the chamber. As such, the contaminants may be chemically distinct from contaminants in Chamber 1 and are likely in higher concentration.

\subsection{Material Inventory of Vacuum Chamber}
Here we provide an inventory of what are (to our knowledge) all the materials placed in Chamber 1 at any point. While we do not know the source of LIC precursors, it is either due to one of these materials (in particular the non-metals) or some trace contaminants introduced during chamber assembly. All materials that were not received in UHV packaging from the manufacturer were solvent cleaned, using standard ultrasonic cleaning UHV procedures wherever possible.

\begin{itemize}
    \item Metals: Copper, Stainless Steel, Gold, Beryllium-Copper, Holmium, Titanium, Platinum, Tungsten, Brass
    \item Valves: VAT and Vacgen All-metal leak valves, VAT UHV Gate valves (contains Viton)
    \item Proprietary: ColdEdge custom-designed UHV/cryogenic compatible sample heater and high-temperature interface mount integrated into a Sumitomo RDK415D-2 cold head
    \item Temperature sensors: Lakeshore DT-670B SD package, Pt-111, and Rhodium-Iron sensors
    \item Miscellaneous insulation, and mounting: Macor, PEEK (Accuglass), Low outgassing epoxy (Masterbond EP21TCHT-1)
    \item Wiring: Lakeshore cryogenic wire  (WQT-36, WMW-32), Accuglass UHV Coaxial cable (Part 110755), Delft Cryoflex 1
\end{itemize}
\section{Additional LIC Measurements}
\subsection{Comparison of LIC under oxygen and nitrogen}
\begin{figure}[h]
    \centering
    \includegraphics[width=\textwidth]{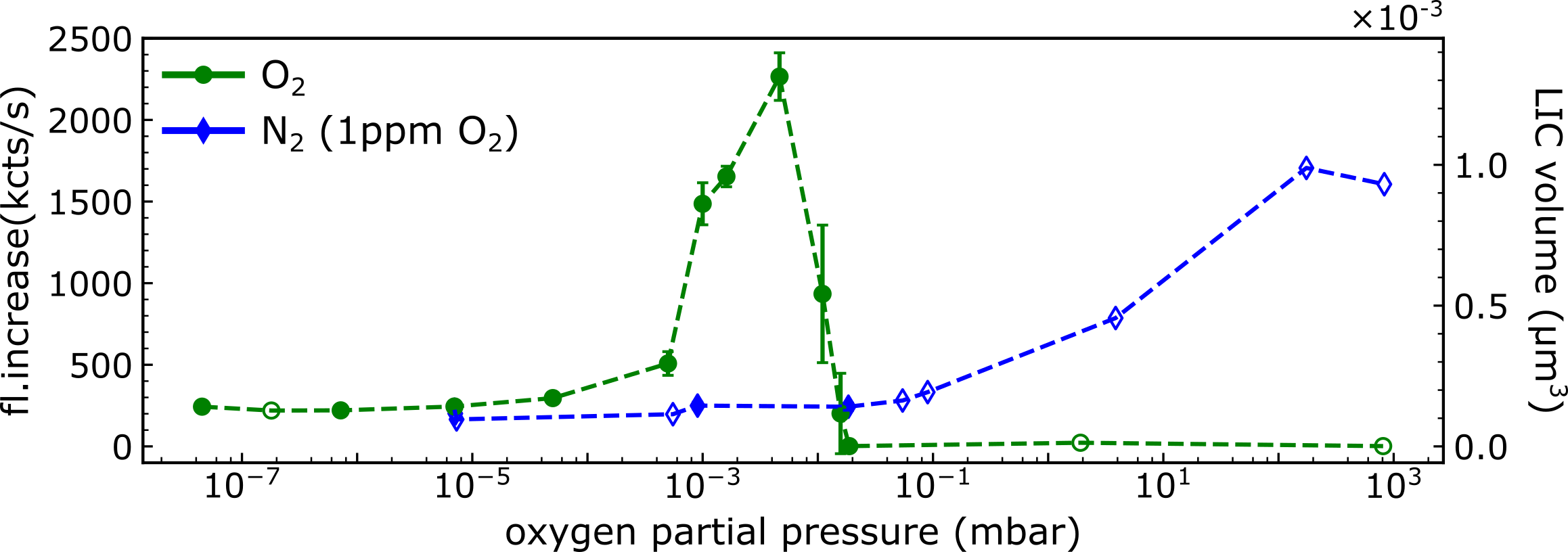}
    \caption{Plotted in green circles is the same \ch{O2} data as the main paper. Plotted in blue diamonds are the results from performing the same experiment described in the main text, but using leaked \ch{N2} as the dominant gas instead.}
    \label{supfig:n2}
\end{figure}
In order to isolate the effect of oxygen from the effect of total pressure, we did a separate experiment leaking nitrogen into the chamber. The oxygen concentration in the leaked nitrogen (as specified by Airgas) was 1ppm \ch{O2}, though it was not measured via any other method. The entire trend is shifted by a factor between $10^4$ to $10^6$, which is consistent with oxygen concentration being the causal factor in determining LIC formation.

We do note that the dependence appears less sharp in Fig. \ref{supfig:n2}, which suggests that total pressure, \ch{N2}, or some other correlated variable (e.g. \ch{H2O} pressure) may have higher-order effects on LIC kinetics. Four possible factors contributing to the slightly lower and broader trend are:
\begin{itemize}
    \item In the context of the adsorbate-based model presented in the main text, an increase in \ch{N2} may displace oxygen and adsorbed LIC precursor molecules. To the extent that the single-layer adsorption model holds, we would in general expect a lower growth rate due to Reaction $A$
    \item Alternatively, the high total pressure dominated by \ch{N2} may quench or otherwise interact with photo-excited oxygen or LIC precursors, reducing the concentration or effective interaction cross-section of excited reactants available for LIC growth.
    \item Concentrations of hydrocarbons in the \ch{N2} gas are specified to ppm levels as well. While we don't have reason to believe that those particular hydrocarbon contaminants are the kind that can produce LIC, that is also a possible confounding factor.
    \item We have observed that an increase in the relative humidity of the chamber slows down the rate of LIC growth, suggesting water may be an additional source of reactive oxygen under the right circumstances. Water levels in this experiment are typically a fixed fraction of the total pressure regardless of the gas (since the lines used for gas introduction were unbaked). Therefore, the total amount of reactive oxygen due to residual water vapor may be higher at the peak of the nitrogen curve relative to the peak of the oxygen curve. This would cause a corresponding decrease in LIC growth and potentially complicated adsorbate dynamics.
\end{itemize}  

Regardless of these perturbative effects, the conclusion is clear: molecular oxygen (and not the total pressure itself) appears to be the causal mechanism for the changes in LIC growth rate detailed in the main text.
\subsection{LIC on silicon}
We placed doped $p$-type silicon, undoped silicon, and our lightly nitrogen-doped diamond substrates in Chamber 2 (see Section \ref{sec:chambers}) during a single pump-down, and studied the laser induced contamination rate under constant power illumination. Exact parameters for each substrate were:
\begin{itemize}
    \item Lightly nitrogen-doped diamond: PECVD-grown diamond on electronic grade substrate. Delta-doped with nitrogen so the resulting \SI{6}{nm} thick doped layer resides \SI{81}{nm} from the surface with a peak density of $4\times 10^{17}\text{atoms}/\mathrm{cm^3}$. Mean square surface roughness of \SI{470}{pm}.
    \item Undoped silicon (University Wafer 3328): Intrinsic (100) silicon. Resistivity $20\mathrm{k\Omega \cdot}$cm, SSP
    \item $p$-type doped silicon (University Wafer 2631): Boron doped (100) silicon. Resistivity $0.001-0.005 \mathrm{\Omega\cdot}$cm, DSP
\end{itemize}
\begin{figure}[h]
    \centering
    \includegraphics[width=0.8\textwidth]{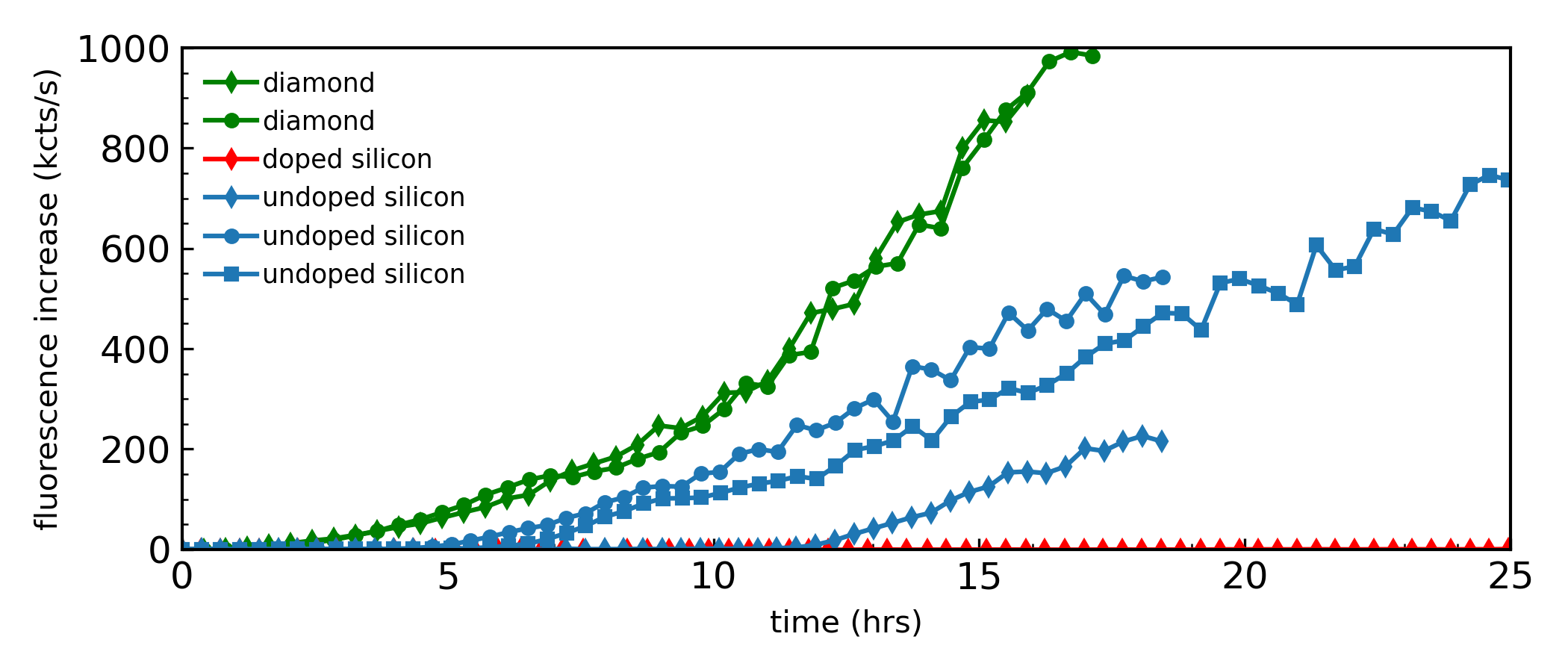}
    \caption{Increase in fluorescence (attributed to LIC) over hours-scale illumination time ($\sim$mW laser power) on 3 different bulk substrates exposed to the same residual vacuum environment. LIC onset and rate following onset were fastest on bulk diamond, but it was still reproducible on undoped silicon substrates. Onset did not occur for the one attempt at producing LIC on a doped silicon substrate. All data were filtered with a moving average and small dips and oscillations are likely not physical, but due to small laser spot drifts during the measurement.}
    \label{fig:silicon}
\end{figure}

From Fig. \ref{fig:silicon}, we see that insulating silicon is susceptible to LIC as well. The LIC growth appears to start later in time relative to insulating diamond, but has a similar functional form. The doped silicon could not be contaminated over more than 25 hours. If repeatable, these data suggest that the formation chemistry for LIC depends critically on charge dynamics in the substrate.
\clearpage
\subsection{LIC vs laser power}
Here we present a cursory study of LIC onset at an oxygen pressure of $1.1\times 10^{-3}$ mbar on single nanopillars of diameter $\sim$\SI{450}{nm} on an otherwise identical sample to the one used in the main text. Only one nanopillar was contaminated at each laser power, and the absolute growth rates should not be compared to those extracted in the main text due to the different nanopillar diameter and that the partial pressure of water in the chamber was likely 2-3 orders of magnitude higher due to preceding experiments.

\begin{figure}[h]
    \centering
    \includegraphics{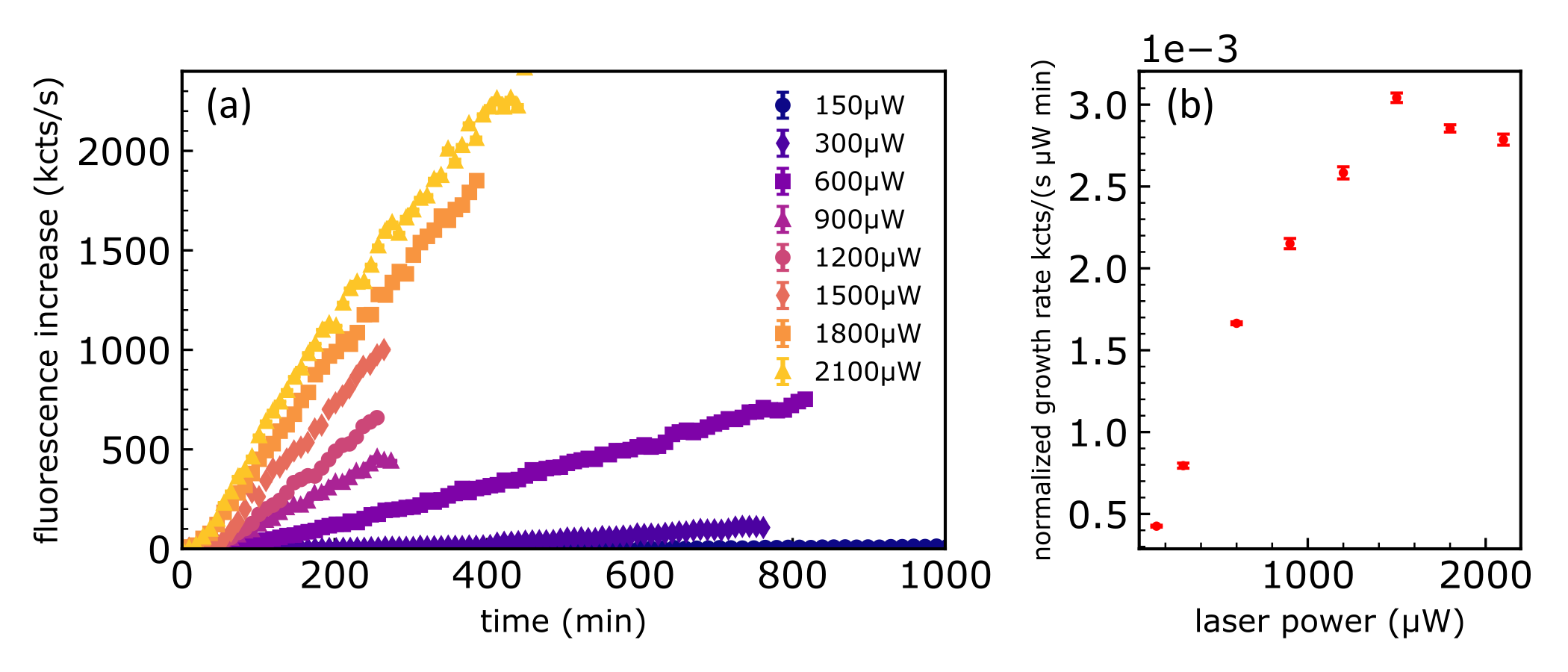}
    \caption{(a) Fluorescence growth vs time for various laser powers, along with piece-wise linear fits of the data. (b) Extracted contamination rate after LIC onset vs incident laser power. Note the contamination rate is in units of (kcts/s/$\mu$W)/min because the excitation power used to probe fluorescence was increased as well, so the rates had to be normalized for comparison.}
    \label{supfig:power}
\end{figure}

In Fig. \ref{supfig:power}(a), we show the fluorescence growth data collected along with piece-wise linear fits used to extract effective onset times and growth rates after onset. No clear trend in onset time was observed, but the rate after onset (normalized to incident laser power) is plotted in Fig. \ref{supfig:power}(b). A regime where LIC growth rate saturates is possibly reached.

\clearpage
\subsection{Evidence of carbonaceous LIC}
While the volumes of LIC created in our work are not easily studied via bulk spectroscopic methods (e.g. secondary ion mass spectroscopy), a surface-sensitive and highly spatially-resolved method such as Auger electron spectroscopy (AES) may help exclude possible elemental contaminants that contribute to LIC. Due to the low energy of the Auger electrons which comprise the detected signal, the measured spectrum reflects the elemental composition of only a shallow depth of up to $\sim10$ atomic layers \cite{mathieu_auger_1997,smith_surface_1994}. Therefore, when the incident beam used to produce Auger electrons is focused on a region that includes LIC, the measurement is primarily sensitive to the volume occupied by contamination.

\begin{figure}[h]
    \centering
    \includegraphics[width=0.88\textwidth]{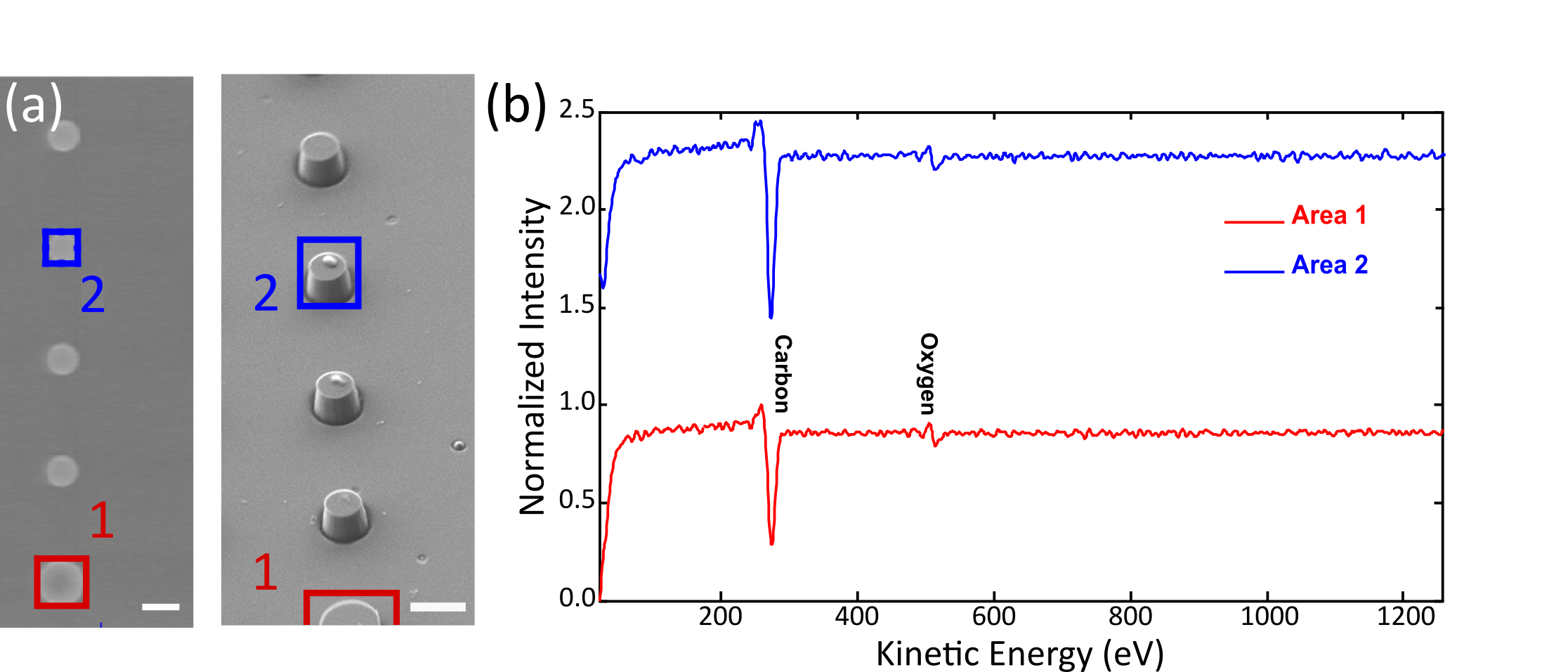}
    \caption{(a) A side-by-side comparison showing the target regions as seen in the SEM imaging mode of the AES instrument using a 10 kV 1 nA beam, and the same pillars imaged in a separate SEM showing the contamination on Area 2. Area 1 is partially cut off, but was known to be uncontaminated. Scale bar is 1$\mathrm{\mu m}$. (b) Smoothed, shifted, normalized, and differentiated Auger spectra for Areas 1 and 2. All data processing was done in the PHI, Inc. MultiPak software. Raw data were smoothed and differentiated with 9-point binomial smoothing and derivative filters. Spectra were normalized to a peak intensity of 1 and horizontally shifted such that the carbon peaks is at 275 eV. The spectrum corresponding to Area 2 is offset vertically for clarity. }
    \label{fig:aes}
\end{figure}

We performed AES (Physical Electronics Auger 700, PHI Inc.) on an uncontaminated (Area 1) and contaminated (Area 2) pillar on the sample. In Fig. \ref{fig:aes}(a), we show the pillars as seen on the AES instrument (left) and as seen using the SEM from the main text (JEOL IT800SHL, right) side-by-side. LIC is not visible on the AES instrument due to the larger voltage and current used, so the entire pillar was included in the target region to ensure the collected AES signal included LIC. In Fig. \ref{fig:aes}(b), we show the differentiated spectrum (see caption for details). The only clear peaks can be attributed to carbon and oxygen. The lack of any other obvious peaks suggests LIC is primarily carbonaceous. 

\clearpage
\section{Physical features of LIC}
\subsection{Deposit location and shape}
\begin{figure}[h]
        \centering
        \includegraphics{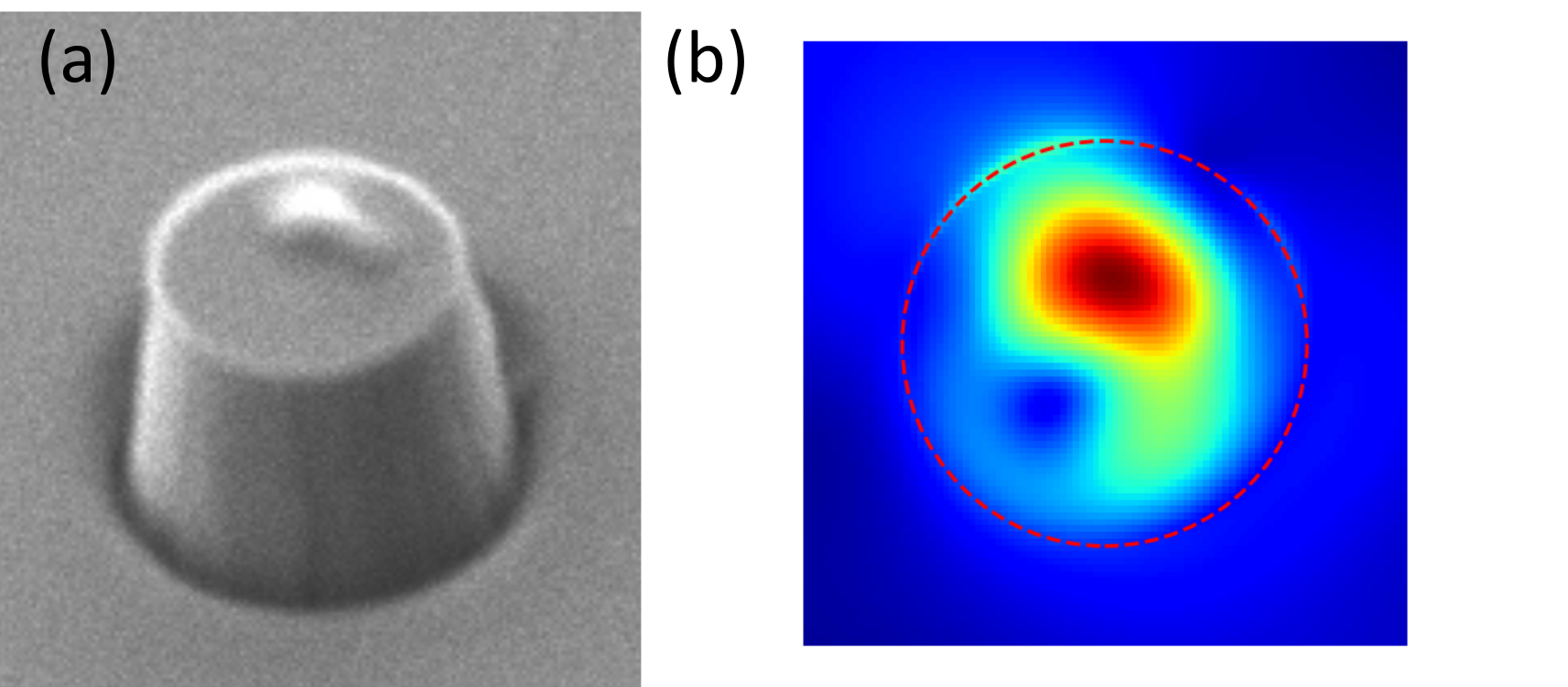}
        \caption{(a) Zoomed in LIC on a nanopillar (b) FDTD Lumerical simulation of average electric field intensity of a 532 nm pulse of green light just above a nanopillar of the same dimensions, with a $10$ degree tilt upon entry. Red dotted outline represents the edge of the pillar. The off-center maximum is due to the beam tilt.}
        \label{fig:lumerical}
    \end{figure}

The off-center location of LIC deposits relative to the nanopillars is likely an indicator that the area of maximum electric field intensity (and thus the area with the highest concentration of photo-activated reactants) is not in the center of the nanopillar. Fig. \ref{fig:lumerical} shows a zoom-in of a contaminated pillar with a Lumerical simulation of the electric field at the surface of a \SI{670}{nm} pillar produced by a gaussian beam focused onto the center of the nanopillar at a 10 degree angle relative to normal incidence. Based on these simulations, we believe a slight misalignment between the sample surface normal and the excitation beam can produce the observed off-center deposit morphology.
    
\subsection{Volume estimation procedure and uncertainty analysis}
\begin{figure}
    \centering
    \includegraphics{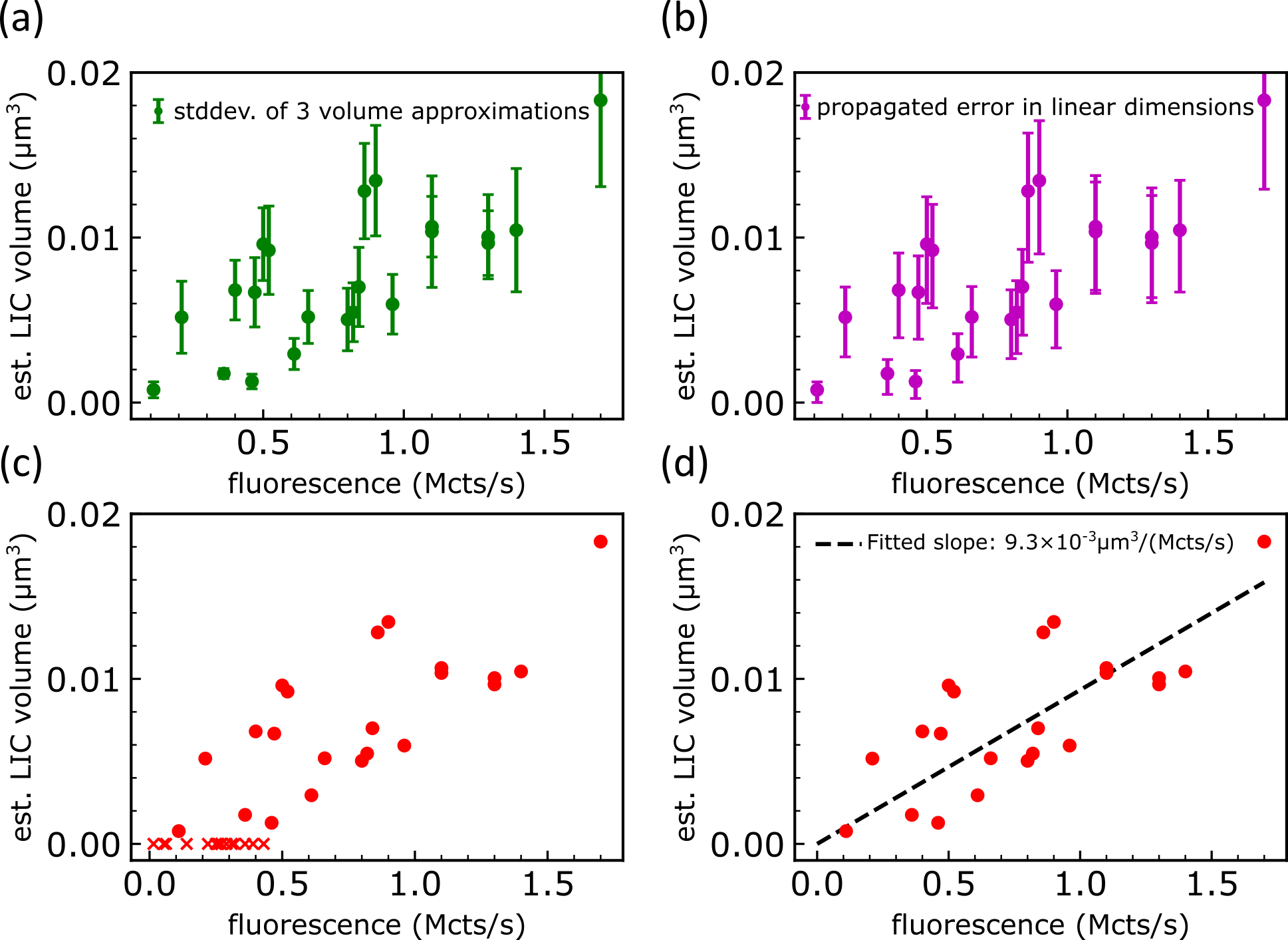}
    \caption{All subfigures show the same underlying scatter plot from the main text, with (a) an estimate of the error based on the choice of approximation for volume (b) propagated error due to the method of length estimation (c) insensitivity of SEM to levels of contamination visible in the fluorescence scan (`x' markers denote pillars that do not obviously have mounds of LIC in the SEM scan) and (d) the best fit line to extract an order-of-magnitude estimate for the volume of LIC created at a particular fluorescence}
    \label{supfig:repeatability}
\end{figure}

Volume estimation of the LIC deposits was performed by visual inspection of the SEM image and comparison to the native scale bar in the SEM software. Since the SEM electron beam probed the sample at a 45 degree angle of incidence, the SEM contrast contains 3D information, i.e. depth $d$, width $w$, and height $h$ of the deposit are all partially discernible from the contrast in the SEM image. After estimation of the three linear dimensions, LIC volume was calculated assuming that the shape of the deposit is closely approximated by a spherical cap where the base of the cap is approximated by the mean of $d$ and $w$, and the height is $h$. We note 2 sources of scatter in our volume estimation procedure and 1 intrinsic source of uncertainty: 
\begin{enumerate}
\item Different reasonable choices for volume shape scale with height and width slightly differently, and have different prefactors. We can approximate the error due to this by calculating the standard deviation of volume estimates produced by 3 different choices for a volume formula: a cylinder ($V = \pi \ell^2 h$ where $\ell = \frac{w + d}{2}$), a gaussian profile ($V = \frac{2 \pi h w d}{2 \times 2.355}$, where we assume the width and depth are the full-width at half maximum in each dimension), and a spherical cap ($V = \frac{\pi}{6} h^3 + \frac{\pi}{8}d^2 h$). Fig. \ref{supfig:repeatability}(a) shows the LIC volume estimates from the main text with this measure of uncertainty.
\item Each linear dimension estimate has some uncertainty. We assume that this measurement uncertainty is nearly independent of the size of the deposit. Based on inspection (varying the length estimates used by a fixed fraction and inspecting how closely they approximate the LIC deposit visually), we choose an uncertainty that is roughly $\sim 10\%$ of the typical estimate in each dimension. Fig. \ref{supfig:repeatability}(b) shows the LIC volume estimates with this fixed absolute error.
\item Lastly, intrinsic to the method used, is the insensitivity of SEM to the smallest volumes of LIC that are discernible in fluorescence scans. We estimate the sensitivity by including all pillars with known LIC-induced fluorescence intensity, even if the LIC was not clearly measurable in the SEM images taken. Fig. \ref{supfig:repeatability}(c) shows these pillars (denoted with `x' markers) overlaid onto the data from the main text. As can be seen by the scatter along the $x$-axis, fluorescence measurements are more sensitive to LIC volume than SEM at the resolution used. Volume error in the SEM image of $\pm$ 0.003 $\mu \mathrm{m}^3$ seems consistent with these data. We believe we could be insensitive to such amounts of LIC spread over the pillar area without further increasing the scan resolution around each pillar individually. 

Since one of our goals with this measurement is to use fluorescence measurements to estimate smaller LIC volumes than are visible on the SEM image, we only base our extraction of a volume-to-fluorescence ratio on the data where contaminants are clearly visible and a higher-quality estimate of volume can be made (fit shown in Fig. \ref{supfig:repeatability}(d)), and then we assume this linear trend holds for lower fluorescence intensities as well. 
\end{enumerate}

To arrive at the quoted volume-to-fluorescence ratio, we need to account for the lower laser power (92 $\mu$W) used to collect the fluorescence data for the scatter plot in Fig. \ref{supfig:repeatability}(d). The fitted ratio is proportionally adjusted so that it can be used for conversion of fluorescence to volume when using the 1.48 mW excitation power used in Fig. 3 of the main text: $9.3\times10^{-3} \mathrm{\mu m^3 \ s/Mcts} \times \frac{0.092}{1.48} \approx 6 \times 10^{-4} \mathrm{\mu m^3 \ s/Mcts}$, the ratio used in the main text.

As can be seen from this analysis, the positive correlation between LIC and fluorescence persists under reasonable error assumptions, and much of the observed scatter may be explained by these errors. Importantly, we therefore do not attempt to glean any additional physical information (e.g. estimates of the variance in LIC volume from pillar to pillar) from these SEM measurements. They are used purely to establish the \textit{order-of-magnitude} estimates of the LIC volumes we create in this study.

\section{Models for LIC growth}
In this section we discuss in more detail the model for contamination presented in the main text. The role of oxygen in photo-chemical and/or auto-catalytic reactions is known to be extremely complex \cite{bhanu_role_1991,kk_rohatgi-mukherjee_fundamentals_1986}, and is itself a field of study. Without detailed knowledge of the chemical precursors of LIC, it is difficult to theorize exact reaction mechanisms (as is done in e.g. Refs. \cite{bhanu_role_1991, delzenne_photosensitized_1964}). Instead, we opt for an effective rate-based model (similar in spirit to what is done in Ref. \cite{stewart_absolute_1989}). 

To specify kinetics for the model in Section \ref{sec:basicmodel}, we adopt some language and inspiration from photo-polymerization. We do not attempt to precisely identify photo-sensitizers, routes for energy or electron transfer, or to assert the intermediate species conjectured here are entirely sufficient. These details are extremely important as they may modify the predicted reaction orders or quantum yields. As is discussed in Section \ref{sec:expandedmodel}, they may even be the fundamental driver of some of the regime changes that we currently attribute to adsorbate effects. Still, our goal with the less precise model presented here is to provide a starting point for photochemical modeling of LIC formation and to emphasize that the measurements in the main text do heavily constrain the possible rate constants and species involved in reactions occurring at the diamond surface. 

Future experiments of the kind done in this work can be designed to rule out or fill in the theoretical details and should directly lead to a more fundamental understanding of LIC formation. Most importantly for NV measurements, such an understanding should also reveal the quantitative levels and reactivities of intermediate species that are available and likely to participate in other destructive photochemistry at the diamond surface. 

\subsection{Adsorbate-based model from main text}
\label{sec:basicmodel}
Our effective description involves the following components:

\begin{itemize}
    \item \ch{M}: a ``monomer" species. The contaminant molecules that serve as precursors to LIC
    \item \ch{L^{(n)}} a ``polymer"-like species that results from the reaction of $n$ monomer sub-units (in our case either linked or catalyzed via oxygen)
    \item \ch{O2}: oxygen (either in gaseous or adsorbed form). Here we assume \ch{O2} is physisorbed. Dissociative chemisorption can be modeled by adjusting the surface coverage ratio according to a modified Langmuir isotherm. We did not find this to significantly improve agreement with measurements, and so for simplicity we do not include it.
    \item \ch{S}: the diamond surface, here participating purely as a host for the adsorption sites for \ch{M} and \ch{O2}, and as a binding site for the starting LIC link \ch{L^{(1)}}
\end{itemize}

To incorporate photo-activation without precise knowledge of the LIC formation mechanism, we assume each reaction requires one single-photon excitation event to promote one of the reactants into a reactive (activated) state. Our core assumption is that excitation is much faster than other processes, which means that an increase in light intensity only modifies the steady-state supply of excited species available for further reactions. For example, in reaction $A$ below, $[\ch{M^*_{(ad)}}] = f(I) [\ch{M_{(ad)}}]$, where the dependence on laser intensity $I$  is entirely captured by the factor $0 \leq f(I) < 1$, the steady-state fraction of excited monomers species. Since we are not probing intensity dependence in this study, we absorb this factor into the overall rate coefficient for the reaction in the discussion below.

We assume the lettered processes described in the main text consist of the following reactions:
\begin{enumerate}[A)]
    \item Adsorption and photo-excitation of LIC precursor \ch{M}, followed by a reaction with adsorbed oxygen via the Langmuir-Hinshelwood mechanism to produce LIC bonded to the diamond surface. In this intermediate-species agnostic description, the choice of exciting \ch{M} or \ch{O2} is arbitrary:
    \begin{align*}
    \ch{M_{(ad)}} + h\nu &\rightarrow \ch{M_{(ad)}^*} \\ 
    \ch{M_{(ad)}^*} + \ch{O2_{(ad)}}  + \ch{S} & \rightarrow  \ch{L^{(1)}S},
    \end{align*}
    with a rate of: $$ \Gamma_A = k_A \left[\ch{O2_{(ad)}}\right] \left[\ch{M_{(ad)}}\right]$$
    
    \item Addition of one unit to LIC via direct interaction with gaseous monomer species (independent of oxygen, but still photocatalyzed):
    \begin{align*}
    \ch{L_{(s)}^{(n)}} + h\nu &\rightarrow \ch{L_{(s)}^{(n)*}} \\
    \ch{M_{(g)}}+ \ch{L_{(s)}^{(n)*}} & \rightarrow  \ch{L_{(s)}^{(n+1)}},
    \end{align*}
    with a rate of: $$ \Gamma_B = k_B \left[\ch{M_{(g)}}\right] \left[\ch{L_{(s)}}\right]$$

    \item Etching of LIC (removal of one unit) via reaction with oxygen
    \begin{align*}
    \ch{O2_{(g)}} + h\nu &\rightarrow \ch{O2_{(g)}^*} \\
    \ch{O2_{(g)}^*} + \ch{L_{(s)}^{(n)}}&\rightarrow \ch{L_{(s)}^{(n-1)}}
    \end{align*}
     with a rate of: $$ \Gamma_C = k_C \left[\ch{O2_{(g)}}\right] \left[\ch{L_{(s)}}\right],$$
     where we have included the solid-state LIC reactant in the rate equations above because we believe the rates in reactions $B$ and $C$ are in some regimes limited by available sites on the solid LIC mass.
\end{enumerate}

Each of the rates above can be re-expressed as a function of partial pressure of the gaseous species and LIC volume $V$ as follows. Here $P_i$ is the partial pressure of species $i$, $\theta_i \equiv \frac{K_i P_i}{1 + \sum_j K_j P_j}$ is the surface coverage of species $i$ as is typically defined for a multi-species, non-dissociative Langmuir isotherm, $N_S$ is the number of available surface sites for adsorbates to attach to, and $N_L$ is the number of sites on the LIC surface available for direct reaction with gases:
        \begin{align}
        \Gamma_A(t) &= k_A \left[\ch{O2_{(ad)}}\right] \left[\ch{M_{(ad)}}\right] = k_A (\theta_{\ch{O2}} N_S(t)) (\theta_{\ch{M}} N_S(t)) = k_A (\theta_{\ch{O2}} \theta_{\ch{M}}) N_S(t)^2 \\
        \Gamma_B(t) &= k_B \left[\ch{M_{(g)}}\right] \left[\ch{L_{(s)}}\right] = k_B P_{\ch{M}} N_L(t) \equiv k_B^{'} N_L(t) \\
        \Gamma_C(t) &= k_C \left[\ch{O2_{(g)}}\right] \left[\ch{L_{(s)}}\right] = k_C P_{\ch{O2}} N_L(t)
        \end{align}
Lastly, we take into account the finite size of the laser spot and the depletion of surface sites as LIC volume grows explicitly in the definition of $N_L(t)$. For an LIC volume $V(t)$, we assume $N_L(t)$ scales as $V(t)^{2/3}$ up to a laser-spot-imposed upper bound $N_{\text{laser}}$. The initial value of $N_S(t)$ is also laser-spot bounded and changes in proportion to total cross-sectional diamond area gained or lost as the LIC volume changes.

To summarize, to calculate the mean LIC growth trajectory, we numerically integrate the codependent evolution of LIC volume and available diamond surface sites:
\begin{align*}
    V(t, N_S) &= V(t-\mathrm{d}t) + (\Gamma_A(t, N_S) + \Gamma_B(t,V) - \Gamma_C(t,V)) \mathrm{d}t\\
    N_S(t, V) &= \max \left[N_S(t) - [\Gamma_A(t,N_s)]^{2/3} + [\Gamma_C(t,V)]^{2/3}, 0\right]
\end{align*}
    where $N_S$ is assumed to decrease or increase depending on the the cross-sectional area of new volume and the source of that volume (e.g. for reaction $B$, surface sites are not depleted). This method of keeping track of available sites is an additional approximation. But for the parameters plotted below, it is consistent with more detailed numerics that kept track of the number of surface sites explicitly (via the Gillespie method).
\begin{align*}
    &\Gamma_A(N_S, P_{\ch{O2}}) = k_A \ \theta_{\ch{O2}}(K_{\ch{O2}},K_{\ch{M}}) \ \theta_{\ch{M}}(K_{\ch{O2}},K_{\ch{M}}) \ N_S^2 \\
    &\Gamma_B(V) = k_B^{'} \min \left[V^{2/3}, N_{\text{laser}}\right] \\
    &\Gamma_C(V, P_{\ch{O2}}) = k_C P_{\ch{O2}}\min \left[V^{2/3}, N_{\text{laser}}\right]
\end{align*}
where $k_A$, $k_B^{'}$, $k_C$ are effective rate constants, $K_{\ch{O2}}$, $K_{\ch{M}}$ are equilibrium constants (defined as ratios of rate constants for adsorption and desorption), and $N_{\text{laser}}$, $P_{\ch{M}}$ (hidden in the Langmuir isotherm formula) are additional physical constants for this experiment. 

\begin{figure}[h]
    \centering
    \includegraphics[width=\textwidth]{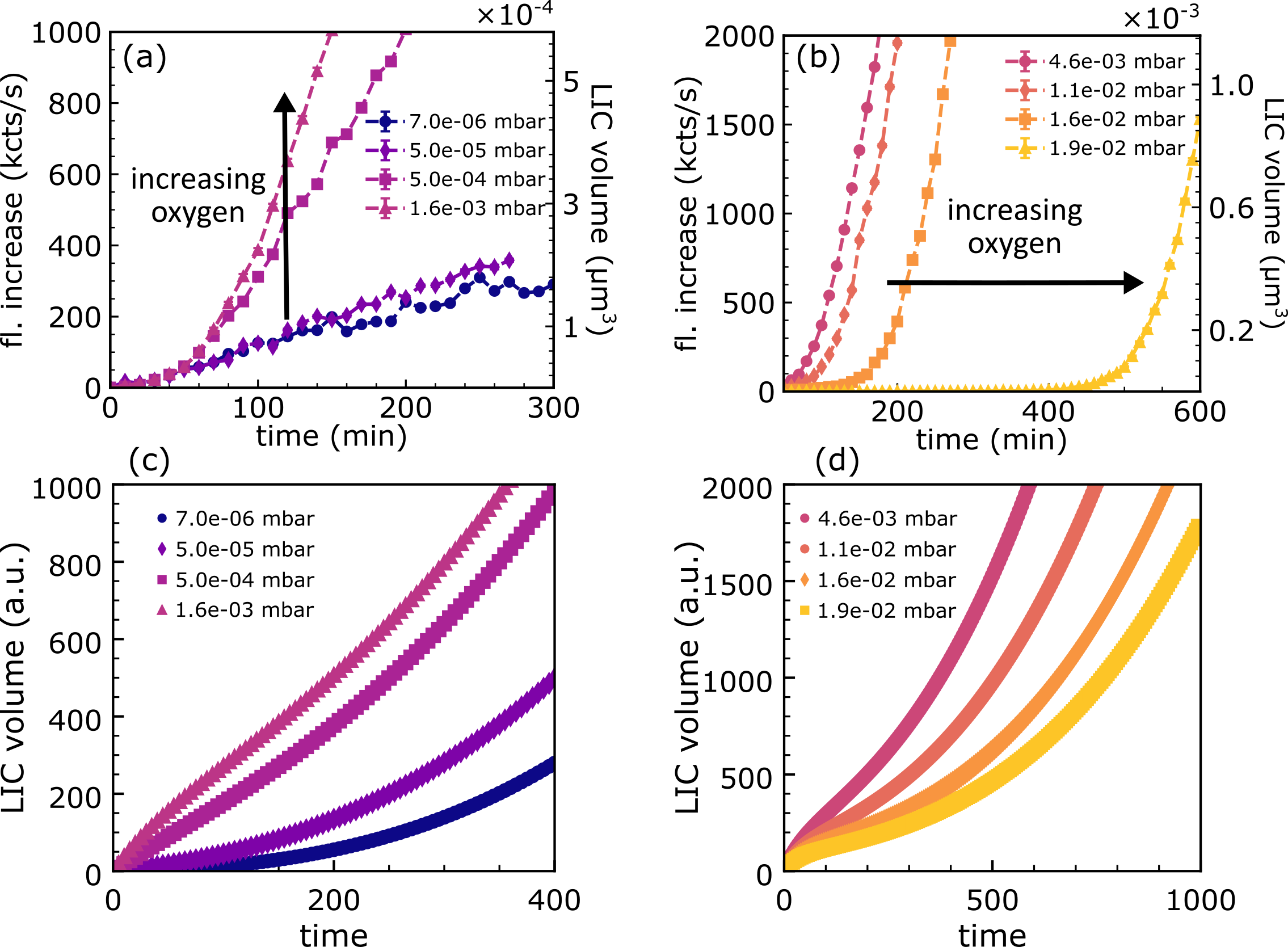}
    \caption{Comparison of model and data. See text for a description of how the ``fit" parameters were chosen. We recover non-monoticity in LIC growth rate and a decay in the onset of the linear regime of LIC growth for larger oxygen partial pressures. The time axis displayed in subfigures (c) and (d) were chosen such that the $y$-axis displayed a similar range to our fluorescence data.}
    \label{fig:model}
\end{figure}

Plotted in Fig. \ref{fig:model} are the data from Fig. 3 of the main text (subfigures (a) and (b)), and simulations corresponding to the same partial pressure of oxygen as the data. These simulations are not a quantitative best fit, but rather a result of applying some heuristic estimates based on physical interpretation of the free parameters. We constrained $K_{\ch{O2}} \gg K_{\ch{M}}$ because hydrocarbon precursors likely have a larger sticking coefficient than molecular oxygen, $P_{{\ch{M}}} < 10^{-9} \mathrm{mbar}$ since no contaminants were obviously detectable with our RGA, and $N_{\text{laser}}$ effectively defined the arbitrary units for the volume axis, chosen to align with units of fluorescence for ease of comparison. Since reaction $A$ is primarily responsible for the non-monoticity, we chose $K_{\ch{O2}}$ by setting the adsorbate product $\theta_{\ch{O2}} \theta_{\ch{M}}$ to be maximized near where we expect the non-monotonicity to occur. Then we set the relative ratios of the rate constants to approximately trade off where we believe the different reactions should dominate. Thus, without over-fitting the many parameters of the model, we are able to recover: (1) Non-monotonic increase and decrease of LIC growth rate over a similar pressure range, (2) A delay in onset of LIC growth as $P_{\ch{O2}}$ is increased, and (3) A complete inhibition of LIC growth at high $P_{\ch{O2}}$ (not shown, as it is evident from the scaling of $\Gamma_C$). We leave quantitative modeling efforts for further work after more aspects of LIC formation are constrained by experiment. 
\label{sup:model}

\subsection{Connection to polymer reactions under oxygen}
\label{sec:expandedmodel}
One feature of the data not reflected in the model is that the late time behavior (at e.g. $P = 7 \times 10^{-6}$ mbar) does not ever appear to transition to a fast growth regime for all low pressures, whereas the model predicts that growth in this regime should do so eventually. It is possible that we did not observe LIC growth for long enough at these pressures and never entered this regime. However, the lack of that behavior may point to different adsorbate dynamics than we assume. 

Another feature of the data is that the transition to fast linear growth should be more sudden than the gradual onset we see in the model simulation. One intriguing approach to address both of these discrepancies is by turning to the well-established role of oxygen as an inhibitor, propagator, and terminator all at once in a wide range of polymerization reactions, from low temperature trialkyl borane-catalyzed polymerization to high-pressure ethylene polymerization \cite{bhanu_role_1991,tatsukami_reaction_1980}. Two features are relevant to LIC. First, Ref. \cite{hansen_kinetics_1964} observes that even for polymerization reactions controlled primarily by oxygen, slow polymerization occurs in the absence of any oxygen. While oxygen may be responsible for our increased rate of LIC growth, it is possible that the baseline growth itself proceeds via a slow, oxygen-independent mechanism in an oxygen-depleted environment. This feature would immediately address the first discrepancy of our model. Second, in the framework established in the references above, oxygen participates in \textit{multiple intermediate steps} of each reaction, reacting with many different radical species at each step. Depending on the precise radical species and their relative concentrations, some of these reactions may leads to products that may not be able to further polymerize (``termination"), may lead to further polymerization (``propagation"), or may react to form the original precursor molecules (``inhibition") \cite{asua_polymer_2008}. This phenomenon allows for non-monotonic behavior \textit{without invoking any adsorbate dynamics} \cite{hansen_kinetics_1964}. For a suitable choice of reaction pathways, this model may display more sudden onset as certain intermediate species are made available.

Thus, there are promising methods in the literature with which our model can be extended  (or an entirely new model constructed) to better account for the data, all of which involve oxygen as the critical ingredient. Further work that may narrow down these pathways may involve forms of desorption spectroscopy \cite{farneth_mechanism_1985} at intermediate points in LIC formation to isolate species (some of which are already used to understand diamond surfaces \cite{thomas_thermal_1992}), more precise fluorescence spectroscopy to tease out molecular structure \cite{kk_rohatgi-mukherjee_fundamentals_1986}, introducing known contaminant molecules \cite{egges_laser-induced_2016}, and further measurement of LIC growth rate as a function of substrate doping.

\clearpage
\bibliography{LIC_Paper.bib}